\documentclass{elsart4-1}

\usepackage{amssymb}

\usepackage[english,francais]{babel}

\usepackage{graphicx}
\IfFileExists{url.sty}{\usepackage{url}}
                      {\newcommand{\url}{\texttt}}

\newtheorem{e-proposition}[theorem]{Proposition}

\newtheorem{e-definition}[theorem]{Definition\rm}

\renewcommand{\theequation}{\arabic{equation}}
\def\og{\leavevmode\raise.3ex\hbox{$\scriptscriptstyle\langle\!\langle$~}}
\def\fg{\leavevmode\raise.3ex\hbox{~$\!\scriptscriptstyle\,\rangle\!\rangle$}}

\begin{document}
\centerline{Neutrino Physics/Physique des neutrinos}
\begin{frontmatter}


\selectlanguage{english}
\title{Reactor Neutrinos}


\selectlanguage{english}
\author[authorlabel1]{Thierry Lasserre},
\ead{tlasserr@cea.fr}
\author[authorlabel2]{Henry W. Sobel}
\ead{hsobel@uci.edu}

\address[authorlabel1]{DAPNIA/SPP, CEA/Saclay, 91191 Gif-sur-Yvette, France \& 
Laboratoire Astroparticule et Cosmologie (APC), 11, place Marcelin Berthelot, 75231 Paris cedex 05, France}
\address[authorlabel2]{Department of Physics and Astronomy, 4129 Frederick Reines Hall, 
University of California, Irvine, CA 92697-4575, USA}
\begin{abstract}

We review the status and the results of reactor neutrino experiments,
that toe the cutting edge of neutrino research. Short baseline experiments
have provided the measurement of the reactor neutrino spectrum, and
are still searching for important phenomena such as the neutrino
magnetic moment. They could open the door to the measurement of coherent
neutrino scattering in a near future. Middle and long baseline oscillation
experiments at Chooz and KamLAND have played a relevant role in neutrino
oscillation physics in the last years. It is now widely accepted that
a new middle baseline disappearance reactor neutrino experiment with
multiple detectors could provide a clean measurement of the last undetermined
neutrino mixing angle~$\theta_{13}$. 
We conclude by opening on possible use of neutrinos for Society~: Non
Proliferation of Nuclear materials and Geophysics.
{\it To cite this article: Th. Lasserre, H.W. Sobel, C. R.
Physique 6 (2005).}

\vskip 0.5\baselineskip

\selectlanguage{francais}
\noindent{\bf R\'esum\'e}
\vskip 0.5\baselineskip
\noindent
{\bf Neutrinos aupr\`es des r\'eacteurs. }

Les exp\'eriences de d\'etection des (anti)neutrinos issus des r\'eacteurs nucl\'eaires
ont jou\'e un r\^ole d\'eterminant en physique des neutrinos depuis 
leur d\'ecouverte. 
Les exp\'eriences au plus pr\`es du c\oe ur nucl\'eaire
ont permis non seulement de comprendre la source
d'antineutrinos, mais \'egalement de contraindre certaines
propri\'et\'es importantes comme le moment magn\'etique du neutrino. 
Une exp\'erience tr\`es proche du c\oe ur pourrait mettre
en \'evidence la diffusion coh\'erente des neutrinos. 
Les exp\'eriences \`a moyenne ou longue port\'ees, Chooz et KamLAND, 
ont jou\'e ces derni\`eres ann\'ees un r\^ole fondamental dans la compr\'ehension
de l'oscillations des neutrinos. De nouvelles exp\'eriences sont en pr\'eparation
pour mesurer le dernier angle de m\'elange encore inconnu, $\theta_{13}$,
en utilisant des techniques similaires mais plus pr\'ecises. Nous concluons
en d\'ecrivant deux applications possibles~: l'utilisation de
d\'etecteurs  d'antineutrinos
pour contribuer \`a la non prolif\'eration des mati\`eres nucl\'eaires fissiles,
et l'\'etude de la g\'eophysique \`a l'aide des antineutrinos ``terrestres''.
{\it Pour citer cet article~: Th. Lasserre, H.W. Sobel, C. R.
Physique 6 (2005).}

 \vskip 0.5\baselineskip
\noindent{\small{\it Mots-cl\'es~:} Neutrinos~; R\'eacteur nucl\'eaire~;
Mot-cl\'e3}}
\end{abstract}
\end{frontmatter}

\selectlanguage{english}

\section{Neutrino discovery}

Invented by Pauli~\cite{Pauli} in 1930, named by Fermi in 1934 and
later modeled in his theory of beta decay~\cite{Fermi}, the extraordinarily
weakly coupling neutrino was first searched for by Reines and Cowan.
Starting at the Hanford nuclear reactor (Washington) they later moved
to the new Savannah River Plant (South Carolina) to perform their
definitive and ground-breaking experimental detection. This feat had
two consequences: resolving and clarifying the highly unsatisfactory
situation of a fundamental particle needed for the consistency of
theory, but virtually unobservable, and demonstrating the possibility
of doing \char`\"{}neutrino physics\char`\"{}. This opened the door
to the use of neutrinos as a sensitive probe of particle physics.
Indeed, several years after the completion of the seminal work of
Reines and Cowan, neutrinos were beginning to be used regularly to
investigate the weak interactions, the structure of nucleons and the
properties of their constituent quarks.

In the first crude experiment of 1953~\cite{Reines1953}, their goal
was to demonstrate unambiguously a reaction caused in a target by
a neutrino produced elsewhere. The experiment pioneered the delayed
coincidence technique to search for the reaction:~$\bar{\nu_{e}}+p\rightarrow e^{+}+n$
where an electron antineutrino from the Hanford nuclear reactor interacted
with a free proton in a large tank filled with cadmium loaded liquid
scintillator. The positron and the resultant annihilation gamma-rays
are detected as a prompt signal while the neutron is thermalized in
the liquid scintillator and subsequently captured by the cadmium.
The excited nucleus then emits gamma radiation which is detected as
the delayed signal. The first result, at two standard deviations,
was followed in 1956 and 1958 by more precise experiments~\cite{Reines1956,Reines1960,Reines1959}
where the significance improved to over four standard deviations.
In addition to the detection, the reaction cross-section was measured
to be~$11\pm2.6\times10^{-44}\textrm{ }cm^{2}$~\cite{Reines1959}. Nowadays, reactor
neutrinos are still detected through similar experimental methods.

\section{Neutrinos from reactors}

\begin{figure}[h]
\begin{center}
\textbf{}\begin{tabular}{cc}
\includegraphics[%
  scale=0.13]{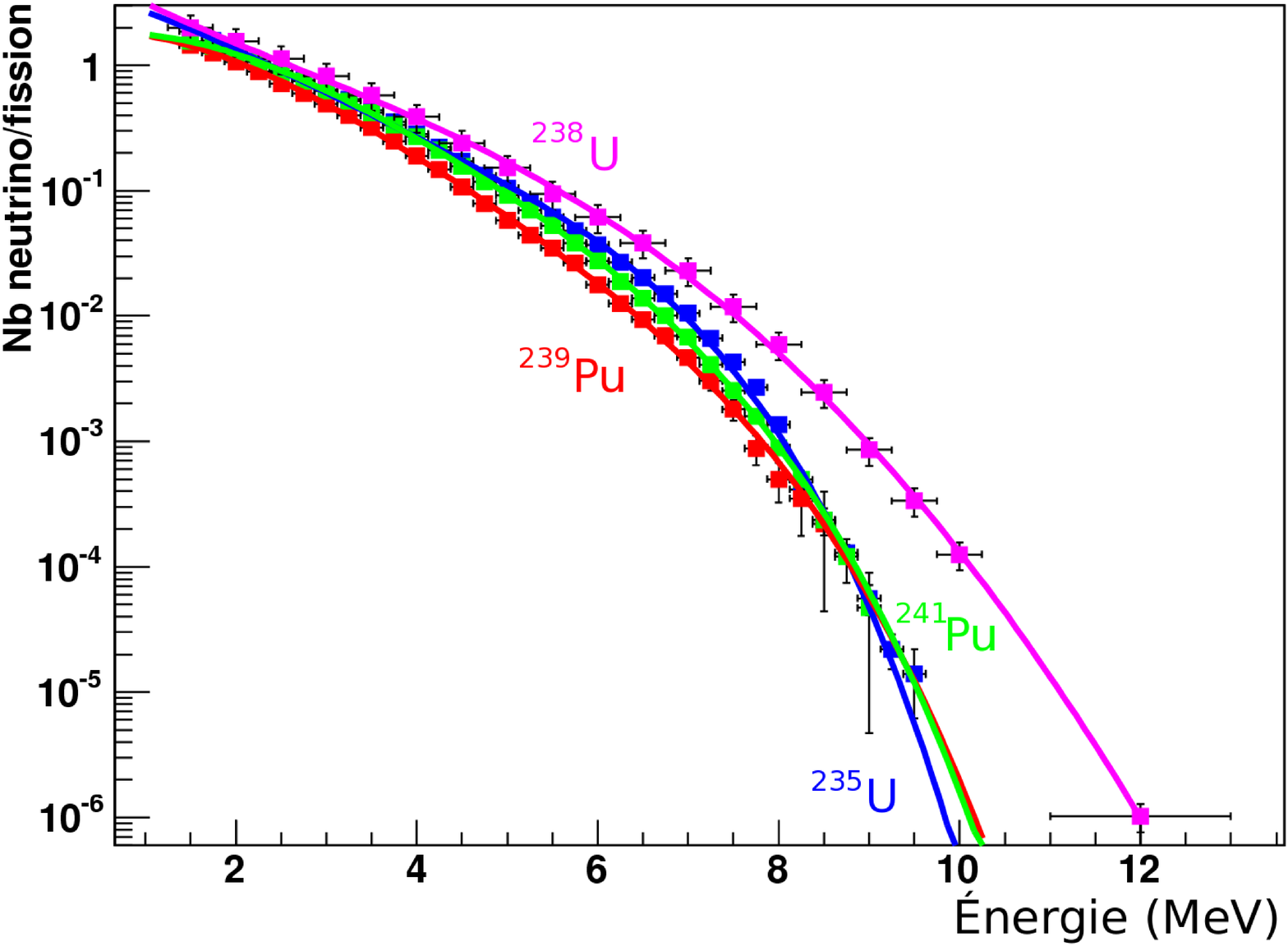}&
\includegraphics[%
  scale=0.27,
  angle=-90,
  origin=rB]{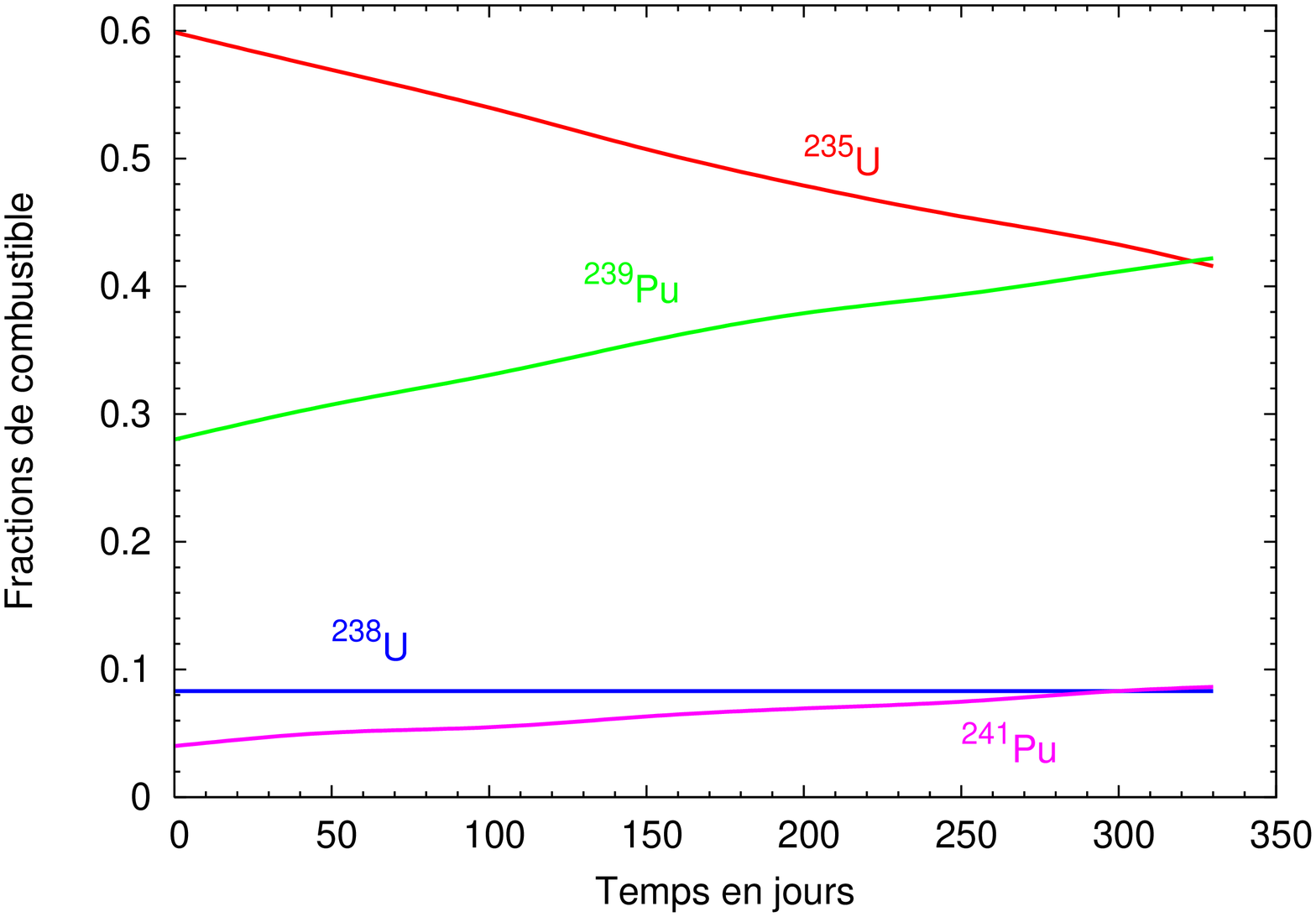}\tabularnewline
\end{tabular}\end{center}

\caption{\label{AntinueSpectra} $\bar{\nu_{e}}$spectra of the four dominant
fissioning isotopes (left). Percentage of fissions of the main fissile
elements during a fuel cycle (right).}
\end{figure}

Fission reactors are prodigious producers of neutrinos~(about $10^{20}$
$\bar{\nu_{e}}$ $s^{-1}$~per nuclear core). The fissioning of~$^{235}$U
produces elements which must shed neutrons to approach the line of
stability. The beta decays of this excess produce approximately six
electron antineutrinos per fission. In modern reactors, the uranium
fuel is enriched to a few percent in~$^{235}$U, but there are also
significant contributions to the neutrino flux from the fissioning
of $^{238}$U, $^{239}$Pu, and $^{241}$Pu. During a typical fuel
cycle, the Pu concentrations increase so the neutrino flux from~$^{239}$Pu,
and~$^{241}$Pu grows with time~(see Figure~\ref{AntinueSpectra}).
The $\bar{\nu_{e}}$~spectrum is calculated from measurements of
the beta decay spectra of~$^{235}$U, $^{239}$Pu, and~$^{241}$Pu~\cite{Schreckenbar}
after fissioning by thermal neutrons. Since~$^{238}$U fissions with
fast neutrons this technique cannot be used. In this case, a summation
of the $\bar{\nu_{e}}$~from all possible beta decay processes is
performed. Since $^{238}$U contributes only about about 11~\% to
the neutrino signal, and further since the error associated with this
summation method is less than 10~\% it contributes less than 1~\%
to the overall uncertainty in the~$\bar{\nu_{e}}$~flux. Experimentally,
the observed rate of positron production from~$\bar{\nu_{e}}+p\rightarrow e^{+}+n$
has been compared to the predicted rate in order to test the precision
of the~$\bar{\nu_{e}}$~spectra prediction~\cite{Achkar}. 

\section{Elastic scattering}

In the Standard Model, the differential cross section for the purely
leptonic (weak scattering) elastic scattering reaction, $\nu_{e}+e\rightarrow\nu_{e}+e$,
is given by\,: \[
\left(\frac{d\sigma}{dT}\right)_{\mbox{\small{\textit{WK}}}}=C\left[g_{L}^{2}+g_{R}^{2}\:(\:1-\frac{T}{E_{\nu}}\:)^{2}-g_{L}g_{R}\frac{m_{e}T}{E_{\nu}^{2}}\right]\:,\]
where~$C=2G_{F}^{2}m_{e}/\pi$, $g_{L}=\sin^{2}\theta_{W}+1/2$ for
$\nu_{e}$, $g_{L}=\sin^{2}\theta_{W}-1/2$ for $\nu_{\mu}$ and $\nu_{\tau}$,
and $g_{R}=\sin^{2}\theta_{W}$ ($\theta_{W}$ is the Weinberg angle). $T=E_{e}-m_{e}$ is the kinetic energy
of the recoil electrons. The first experiment to detect this reaction~\cite{rei}
took advantage of a well shielded location only 11~meters from the
center of the Savannah River reactor. The intense flux of~$2.2\times10^{13}$~$\nu_{e}$~cm$^{-2}$~sec$^{-1}$
and the regular up/down cycling of the reactor gave a reasonable signal
in a small target and eliminated non-reactor-associated backgrounds.
The experimental cross section was measured with a precision of $\pm$26~\%.
Analysis of the experimental results also produced a crude~($\pm$17~\%)
measurement of~$\theta_{W}$.

\section{Coherent neutrino scattering}

If the neutrino energy is small, the neutrino wavelength can be large
when compared to the size of a nucleus. In this case, the neutrinos
can scatter coherently from the nucleons and the scattering cross-section
would be proportional to the square of the weak nuclear charge ~\cite{coh}\,:\[
\frac{d\sigma}{d(cos\theta)}=\frac{G^{2}}{8\pi}\left[Z(4sin^{2}\theta_{W}-1)+N\right]^{2}E^{2}(1+cos\theta)\:.\]

In 1977, Freedman et. al.~\cite{coh} pointed out that the large
cross-section, particularly important for supernova dynamics, should
be experimentally measured at nuclear reactors. Drukier and Stodolsky~\cite{Drukier}
suggested using cryogenic bolometers to observe the very low energy
nuclear recoil (100's~of eV's for~MeV neutrinos), but that technique
has not been successful up to now.

\section{Neutrino magnetic moment}

In the Standard Model, the massless neutrinos do not have magnetic
moments, but this is not the case if neutrinos do have mass~\cite{lee}.
However, introducing this effect in the model leads to moments of
at least eight orders of magnitude below currently accessible experimental
limits. These limits are derived from reactor~$\bar{\nu}_{e}$'s~\cite{rei,der,texo,munu},
in the range of~$(0.9-4)\times10^{-10}\mu_{B}$, where $\mu_{B}$
is the Bohr magneton. Various astrophysical observations also yield
limits on the neutrino magnetic moment in the range from~$10^{-12}\mu_{B}$
to $4\times10^{-10}\mu_{B}$~\cite{hag}. Therefore, a positive observation
of such large magnetic moments would imply additional physics beyond
the Standard Model. Experimentally, neutrino-electron scattering has
been used in the search for neutrino magnetic moments. If~$\mu_{\nu}\neq0$,
the differential cross section of neutrinon-electron scattering is
an incoherent sum of weak scattering and magnetic scattering~\cite{bea}\,:
\[
\left(\:\frac{d\sigma}{dT}\:\right)_{\mbox{\small{\textit{EM}}}}=\mu_{\nu}^{2}\:\frac{\pi\alpha_{em}^{2}}{m_{e}^{2}}\;\left(\;\frac{1}{T}-\frac{1}{E_{\nu}}\:\right)\:,\]
where $\mu_{\nu}$ is in units of $\mu_{B}$, $E_{\nu}$ is the neutrino
energy, $T=E_{e}-m_{e}$, and~$T$($E_{e}$) is the kinetic (total)
energy of the recoil electrons.

The contribution from electromagnetic scattering increases rapidly
at lower energies, so the signature of a non-zero neutrino magnetic
moment would be an enhancement of the event rate at lower energies.
The current constraint including the latest KamLAND data results in
a limit on the neutrino magnetic moment at 90~\% C.L. of $\mu_{\nu}\leq1.1\times10^{-10}$
$\mu_{B}$ with the limit at $\Delta m^{2}=6.6\times10^{-5}\mbox{eV}^{2}$
and $\tan^{2}\theta=0.48$. This result is comparable to the most
recent magnetic moment limits from reactor neutrino experiments of
$1.3\times10^{-10}$ $\mu_{B}$ (TEXONO)~\cite{texo} and~$1.0\times10^{-10}$
$\mu_{B}$ (MUNU)~\cite{munu}, albeit for neutrinos and not antineutrinos.

\section{Neutrino Oscillations}

There is now convincing evidence for flavor conversion of atmospheric~\cite{KajitaLipari},
solar~\cite{CribierBowles}, reactor and accelerator neutrinos~\cite{SK98,osc2,osc3,sno,KamLAND1,K2K}.
Thus, neutrinos do have masses, and neutrino oscillation is the most
promising scenario to explain the data (see \cite{Bouchez} for details on
the neutrino oscillation mechanism). Reactor neutrino experiments
measure the survival probability~$P(\bar{\nu_{e}}\rightarrow\bar{\nu_{e}})$
of the~$\bar{\nu_{e}}$ emitted by nuclear power stations at a given
distance~(L). This disappearance probability does not depend on the
Dirac CP phase $\delta$. Furthermore, thanks to the combination of
the MeV range neutrino energies~(E) and the short baselines (less
than thousand kilometers) the modification of the oscillation probability
induced by the coherent forward scattering from matter electrons (so-called
matter effect) can be neglected in first approximation. If neutrinos
masses satisfy~$m_{1}<m_{2}<m_{3}$ (so-called ``Normal Hierarchy'',
NH), the survival probability can be written:
\begin{eqnarray}
1-P(\bar{\nu_{e}}\rightarrow\bar{\nu_{e}})=4\sin^{2}\theta_{13}\cos^{2}\theta_{13}\sin^{2}\frac{\Delta{m}_{31}^{2}L}{4E}+\cos^{4}\theta_{13}\sin^{2}(2\theta_{12})\sin^{2}\frac{\Delta{m}_{21}^{2}L}{4E}\label{3oscFormulae}\\
-2\sin^{2}\theta_{13}\cos^{2}\theta_{13}\sin^{2}\theta_{12}\left(\cos\frac{(\Delta{m}_{31}^{2}-\Delta{m}_{21}^{2})L}{2E}-\cos\frac{\Delta{m}_{31}^{2}L}{2E}\right)\:.\end{eqnarray}
 The first two terms of the right side of Equation~\ref{3oscFormulae}
are, respectively, the atmospheric~($\Delta m_{31}^{2}=\Delta m_{atm}^{2}$)
and solar driven~($\Delta m_{21}^{2}=\Delta m_{sol}^{2}$) oscillations,
while the third term is an interference between both contributions~\cite{HLMA}.
An experiment is only sensitive to the values of $\Delta m^{2}$such
that $L>L_{\textrm{osc}}(meter)=2.48E(MeV)/\Delta m^{2}(eV^{2})$.
We notice here that $\theta_{13}$ is the mixing angle that couples the
heaviest neutrino field to the electron field~(NH). If $\Delta m_{sol}^{2}<<\Delta m_{atm}^{2}$
and/or $\theta_{13}$ is small enough, the solar driven and the atmospheric
driven neutrino oscillations decouple. The mixing is then radically
simplified, leading to two neutrino mixing formula: \[
1-P(\bar{\nu_{e}}\rightarrow\bar{\nu_{e}})=\sin^{2}2\theta_{i}\cdot\sin^{2}\left(1.27\frac{\Delta m_{i}^{2}[eV^{2}]L[m]}{E_{\bar{\nu_{e}}}[MeV]}\right)\:.\]
For the reactor neutrino oscillations we can consider two extreme
cases: $\Delta m_{i}^{2}=\Delta m_{21}^{2}$ and~$\theta_{i}\sim\theta_{sol}$
if the baseline considered exceeds a few tens of kilometers, and $\Delta m_{i}^{2}=\Delta m_{31}^{2}$
and $\theta_{i}=\theta_{atm}$ if it is does not exceed a few kilometers.

\subsection{The pioneering experiments}

\begin{center}%
\begin{figure}[h]
\begin{center}
\textbf{}\begin{tabular}{cc}
\includegraphics[%
  height=5.5cm,
  keepaspectratio]{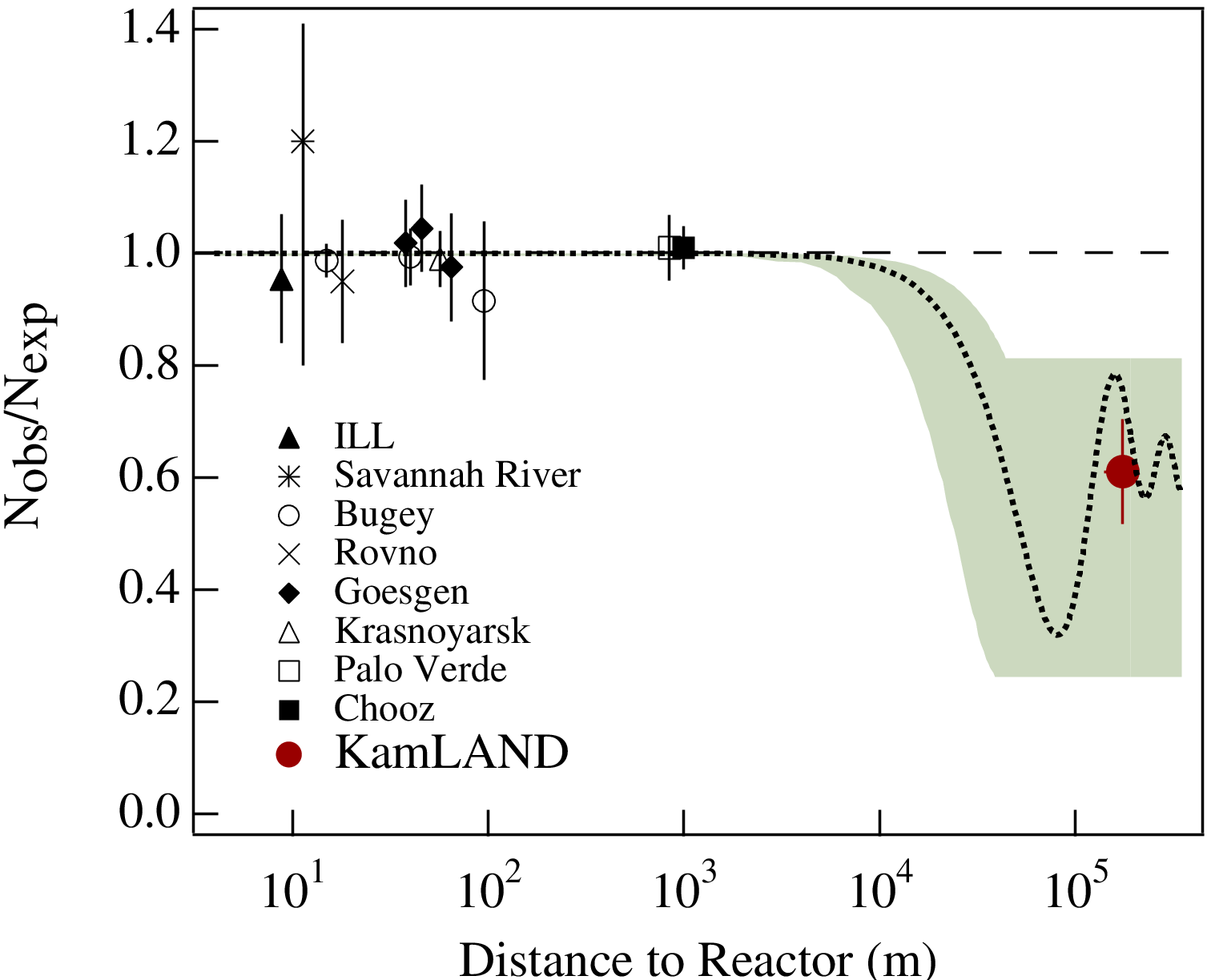}
\includegraphics[%
  height=5.5cm,
  keepaspectratio]{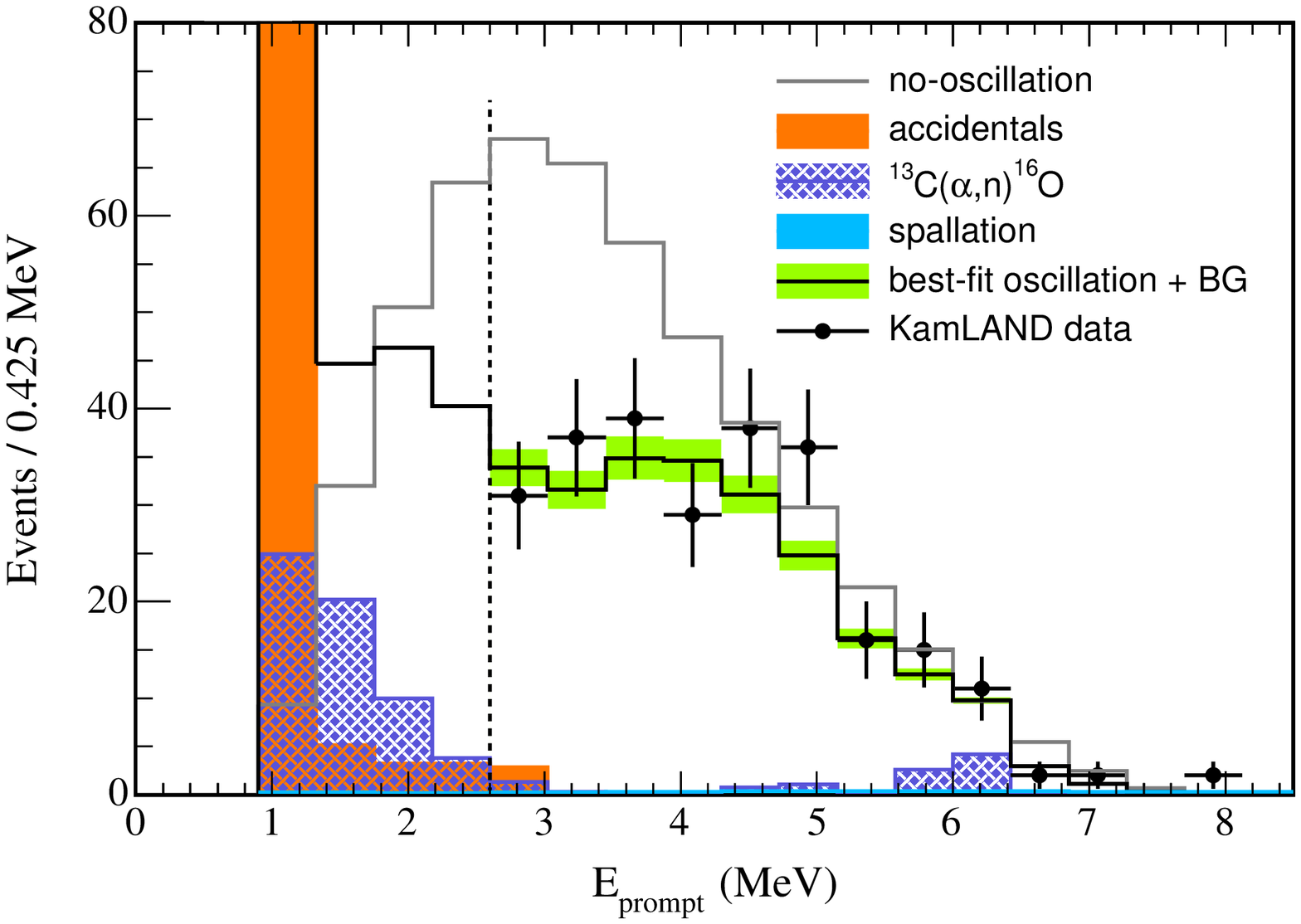}\tabularnewline
\end{tabular}\end{center}

\caption{\label{ReactorExperiments} (left) Ratio of the observed number of neutrino
events with respect to the no-oscillation case for all reactor neutrino
experiments. The source-detector distances are displayed on the x-axis.
Among all experiments, only KamLAND has observed a disappearance thanks
to its very long source-detector average distance~\cite{KamLAND1}.
\label{KamLANDSpectrum} (right) Measurement of neutrino oscillation with
KamLAND, and the evidence of spectral distortion. The energy spectrum
(visible energy in the detector) is displayed together with best-fit
oscillation spectrum (in solid black). The unoscillated spectrum (grey
histogram), and the other backgrounds (accidental, spallation, and
$\alpha$ induced) are also shown~\cite{KamLAND2}.}
\end{figure}
\end{center}

In the eighties and nineties, several experiments~\cite{Goesgen,Rovno,Krasnoyarsk,Bugey}
were performed at a few ten's of meters from nuclear reactor cores
at Goesgen~(Switzerland), Rovno, Krasnoyarsk~(Russia), and ILL Grenoble,
Bugey~(France). Since the knowledge of the neutrino source was not
better than 10~\%, they compared the neutrino rate at different distances
to improve their sensitivity. The most stringent bounds on the oscillation
parameters of this generation of experiments were obtained at Bugey.
The ~$\bar{\nu_{e}}$ spectra were measured at three different source-detector
distances (15, 40, and 95~m), using three identical modules filled
by $^{6}$Li-doped liquid scintillator. Measurements were in agreement
with the no-oscillation expectation, constraining the oscillation
parameters in the region $\Delta m_{atm}^{2}\sim10^{-2}eV^{2}$~\cite{Bugey}.
From this set of experiments (see Figure~\ref{ReactorExperiments}),
the absolute normalization and the spectral shape of reactor~$\bar{\nu_{e}}$
are known to a precision of about 2~\%~\cite{Declais}.

\subsection{Exploring the \char`\"{}atmospheric oscillation\char`\"{}}

In the fall of the nineties, two experiments were performed to test the hypothesis that neutrino oscillations occur
in the parameter region probed by the atmospheric neutrino experiments,~$\Delta m_{atm}^{2}\sim10^{-3}\textrm{ }eV^{2}$~\cite{SK98}. 

The Chooz experiment was located in the Ardennes region of France,
1050~m away from the double unit Chooz nuclear reactors (PWR,
8.4~GW$_{\textrm{th}}$). The detector was located in an underground
laboratory below a 100~m rock overburden (300~m of water equivalent, mwe), providing, for
the first time at reactors, a strong reduction of the cosmic ray induced
backgrounds. The homogeneous detector was filled by a 5 ton Gd-doped
liquid scintillator target, surrounded by a thick active (scintillating)
buffer and a muon veto. The external tank was surrounded by an additional
layer of low radioactive sand. This composition of shielding moderates
neutrons induced by muons outside of the detector as well as the $\gamma$'s
produced by the rocks. Since the two Chooz reactors were commissioned
after the start of the experiment, there was a unique opportunity
to perform an in-situ background measurement. 

The Palo Verde experiment was located in an underground bunker under
12~meters of rock (32~mwe), 750 and 890~meters away from a 3-unit
nuclear power station (11.6~GW$_{\textrm{th}}$) in the Arizona desert.
The low overburden required the use of a segmented detector to reduce
the background. It was composed of 66 acrylic cells of 9~meters filled
with a Gd-doped liquid scintillator, surrounded by a 1~meter thick
water shielding and an efficient liquid scintillator muon veto. 

Neither Chooz nor Palo Verde observed any evidence of neutrino oscillation.
The results could be presented as the energy averaged ratio (R) between
$\bar{\nu_{e}}$ detected and expected $R_{Chooz}=1.01\pm2.8\%\textrm{ }(stat.)\pm2.7\%\textrm{ }(syst.)$
and $R_{PaloVerde}=1.01\pm2.4\%\textrm{ }(stat.)\pm5.1\%\textrm{ }(syst.)$.
Both experiments excluded any $\bar{\nu_{e}}\rightarrow\bar{\nu_{x}}$
oscillation driven by $\Delta m_{atm}^{2}\sim10^{-3}\textrm{ }eV^{2}$,
except for small mixing. Assuming the conservation of CPT, they excluded
the $\nu_{\mu}\rightarrow\nu_{e}$ oscillation solution in the Kamiokande
experiment~\cite{SK98}. The Chooz experiment still provides the
world best constraint on the~$\theta_{13}$ mixing angle\,: $\sin^{2}(2\theta_{13})<0.14$,
at $\Delta m_{atm}^{2}=2.5\textrm{ }10^{-3}\textrm{ }eV^{2}$~\cite{ChoozLast}.

\subsection{Exploring the \char`\"{}solar oscillation\char`\"{}}

A reactor neutrino experiment with a baseline distance of hundreds
of kilometers is sensitive to the Large Mixing Angle~(LMA) oscillation
solution of the solar electron neutrino deficit (see~\cite{CribierBowles}). 
If the reactor-detector
distance is slightly larger than the oscillation length, neutrino
oscillations are observable as an integral reduction of the interaction
rate, as well as a periodic modulation of the~$\bar{\nu_{e}}$ spectrum,
which provides a sensitivity to~$\Delta m_{sol}^{2}$. 

The KamLAND experiment \cite{KamLAND1} is located at the site of the earlier Kamiokande
in the Kamioka mine (Japan), below 2700~mwe of rock. The detector
consists of 1~kton of ultra-pure liquid scintillator contained in
a 13~m diameter transparent nylon balloon suspended in a non-scintillating
oil buffer. The balloon is surrounded by about 1900~PMTs mounted
on a 18~m diameter stainless steel vessel. KamLAND is surrounded
by more than 50~nuclear power units, at an averaged distance of 180~km.
KamLAND has published data from 766 ton-year of exposure. In the absence
of neutrino oscillation 365.2~events are expected (above 2.6~MeV
to eliminate the geo-neutrinos background, see section 7.2), but only 258 $\bar{\nu_{e}}$ candidate
events have been detected. Accounting for 18~expected background
events, the statistical significance for reactor $\bar{\nu_{e}}$
disappearance is 99.998~\%. Assuming CPT invariance, this result
excludes all but the LMA solution to the solar neutrino deficit. It
suggests that solar neutrino flavor transformation through the MSW
matter effect (see \cite{CribierBowles}) has a direct correspondence to antineutrino oscillations
in vacuum. In addition, the energy spectrum measured by KamLAND disagrees
with the expected spectral shape in the absence of neutrino oscillation
at 99.6\% significance and prefers the distortion expected from~$\bar{\nu_{e}}$
oscillation effect (see Figure~\ref{KamLANDSpectrum}). A two-neutrino
oscillation analysis of the KamLAND data gives $\Delta m_{sol}^{2}=\,7.9_{-0.5}^{+0.6}\times$10$^{-5}$~eV$^{2}$~\cite{KamLAND2}.

The Borexino experiment in Italy is designed to detect low-energy
solar neutrinos via electron scattering, using 0.3~kton of ultra-pure
liquid scintillator target, observed by 2200~PMTs \cite{CribierBowles}. 
\textbf{}Borexino is surrounded by many nuclear power
stations, but with a characteristic distance of 800~km (Italy does
not host nuclear power plants). In this case, the shape of the neutrino
spectrum will appear unchanged since the baseline is too large. Thus,
only lower limits on~$\Delta m_{sol}^{2}$ can be derived after a
few years of data taking, since about 30 events are expected each
year in the no-oscillation case~\cite{Borexino}. 

In the future, a new reactor neutrino experiment with a baseline corresponding
to the first oscillation dip (about $60$~km) could provide a high
precision determination of $\sin^{2}\theta_{12}$. With an exposure
of 60~GW$_{\textrm{th}}$kton year and a systematic error of 2~\%,
$\sin^{2}\theta_{12}$ could be determined with an uncertainty of
2~\% at one standard deviation~\cite{Band}.

\subsection{Measuring the third and last neutrino mixing angle $\theta_{13}$}

Considering only the three known families, the neutrino mixing matrix
is parametrized by three mixing angles. The angle~$\theta_{12}$
has been measured to be large, $\sin^{2}(2\theta_{12})\sim0.8$, by
the combination of the solar neutrino experiments and KamLAND
(see \cite{CribierBowles}). The angle~$\theta_{23}$ has
been measured to be close to maximum, $\sin^{2}(2\theta_{23})>0.9$,
by atmospheric neutrino experiments~\cite{SK98,KajitaLipari} as well as the long
baseline accelerator neutrino experiment K2K~\cite{K2K}. However,
we only have an upper limit to the mixing angle~$\theta_{13}$, given
mainly by the Chooz experiment, $\sin^{2}(2\theta_{13})<0.2$. The
large value of both $\theta_{12}$ and $\theta_{23}$ indicates a
strong difference between leptonic and quark mixings, whereas the
smallness of~$\theta_{13}$ testifies to the peculiarity of the neutrino
sector. The value of~$\theta_{13}$ is not only of fundamental interest
to understand leptonic mixing, but it is also necessary to plan for
the future experimental program in neutrino physics, since CP-violating
effects are proportional to~$\sin^{2}\theta_{13}$. 

New accelerator neutrino beams coupled with off-axis detectors, will
search for a~$\nu_{e}$ appearance signal. The observation of a~$\nu_{e}$
excess in an almost pure $\nu_{\mu}$ neutrino beam would be major
evidence for a non-vanishing~$\theta_{13}$. But on the top of the
statistical and systematic uncertainties, correlations and degeneracies
between $\theta_{13}$, $\theta_{12}$, sgn($\Delta m_{31}^{2}$),
and the CP-$\delta$ phase degrade the accessible knowledge on $\theta_{13}$~\cite{Lindner03}.
Both reactor and accelerator programs will provide complementary results
to better constrain the last undetermined parameters~(see \cite{AutieroDeclais}). 

\begin{figure}[h]
\begin{center}\includegraphics[%
  scale=0.55]{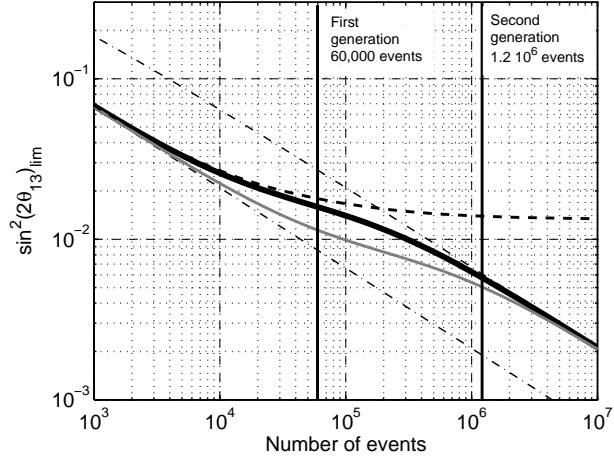}\end{center}

\caption{\label{ReactorTheta13Sens}The sensitivity to $\sin^{2}(2\theta_{13})$,
at the 90~\% confidence level, as a function of the number of events
in the far detector, for two different values of the relative normalization
error: 0.6~\% (thick black curve) and 0.3~\% (thin gray curve).
We considered a far detector located at 1.5~km from the cores, and
a near detector close enough to measure the antineutrinos prior to
their oscillation. We take the atmospheric mass splitting at~$\Delta m_{atm}^{2}=2.5\textrm{ }10^{-3}\textrm{ }eV^{2}$.
The global normalization error is taken to be 2~\%. The vertical
lines mark the luminosities of the first and second generation experiments
(60,000 and 1,200,000 events). The dashed line displays the sensitivity
taking into account a 2~\% uncorrelated background subtraction error,
in both cases; background will thus be crucial for the second generation
of experiments~\cite{Mention05}.}
\end{figure}

In order to improve the Chooz results with reactor experiments, two
(or more) identical detectors close to a power station are required.
The first detector has to be located at a few hundred meters from
the reactor cores to monitor the~$\bar{\nu_{e}}$ flux and spectrum
before the oscillations. The second detector has to be placed between
1 and 2~km away from the core, to search for a departure from the
overall $1/L^{2}$ behavior of the~$\bar{\nu_{e}}$ energy spectrum,
the footprint of oscillation~\cite{WhitePaperReactor}. At Chooz,
the reactor induced systematic error was 1.9~\%, but this class of
uncertainties cancels with the new set up. Two identical detectors
allow relative comparison, leading to a systematic error of 0.6~\%,
using standard technologies~\cite{DoubleChoozEU,DoubleChoozUS}.
Of course, the statistical error has also to be decreased to a similar
amount~\ref{ReactorTheta13Sens}.%

Several sites have been proposed to host a new reactor experiment
to search for~$\theta_{13}$: Angra dos Reis (Brazil), Braidwood,
Diablo Canyon (US), Angra (Brazil), Chooz, Cruas, and Penly (France),
Krasnoyarsk (Russia), Daya Bay (China), and Kashiwazaki (Japan). We
present a selection of the current proposals~\cite{ReactorTheta13Review}.
We classified them into first generation experiments~(see Figure~\ref{ReactorTheta13Sens}),
to be done in the near future, and second generation, more ambitious
projects, which need a significant R\&D effort.

\subsubsection{First generation experiments}

\begin{figure}[h]
\begin{center}
\includegraphics[%
  scale=0.4]{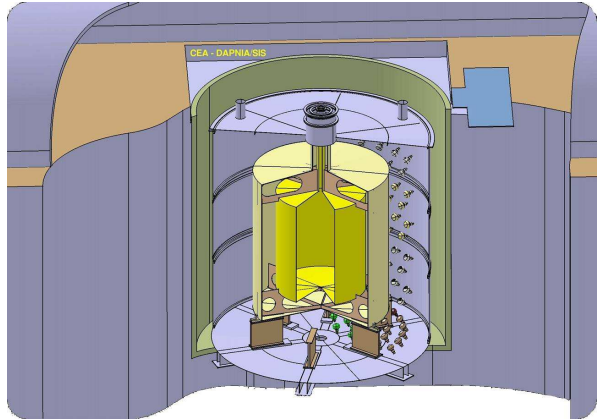}
\end{center}

\caption{\label{DoubleChoozDetector}Overview of the Double Chooz detector
(at the far site, 1.05~km from the nuclear cores).}
\end{figure}

The Double Chooz \cite{DoubleChoozEU} experimental site is located close to the twin reactor
cores of the Chooz nuclear power station, operated by the French company
Electricit\'{e} de France (EDF). The two, almost identical, detectors
contain a 10 ton fiducial volume of liquid scintillator doped with
0.1~\% of Gadolinium (Gd). The underground laboratory of the first
Chooz experiment, located 1.05~km (under 300~mwe) from the
cores is going to be used again. The second detector will be installed
at about 150 m from the nuclear cores. An artificial hill of about 20~m (60~mwe)
height has to be erected above the detector. The detector
design is an evolution of the Chooz detector (see Figure~\ref{DoubleChoozDetector}).
Starting from the center of the target the detector elements are as
follows: the neutrino target\,; a thick acrylic cylinder, filled with
0.1~\% Gd loaded liquid scintillator\,; the $\gamma$-catcher, filled
with unloaded liquid scintillator (the role of this additional region
is to determine the full positron energy, as well as most of the neutron
energy released after neutron capture)\,; a buffer region filled with
non scintillating oil, to decrease the accidental backgrounds from
PMTs radioactivity\,; the stainless steel structure supporting approximately
500 PMTs\,; a muon veto\,; an external shielding of steel protects the
inner detector from the radioactivity of the rock; and finally an
outer muon veto. The dominant error
is the relative normalization between the two detectors. It is expected
to be less than 0.6~\%. Correlated events is the most severe background
source. In total, the background rates (accidental + correlated) for
the near detector would be the order of tens of events for 60~mwe
overburden. For the far detector the total background is estimated
between 1/d and 2/d. This can be compared with the signal rates of
$4,000$/d and $80$/d in the near and far detectors. The expected
sensitivity is: $\sin^{2}(2\theta_{13})<0.025$ (90$~\%$~C.L., for
$\Delta m_{\textrm{atm}}^{2}=2.4\,10^{-3}$~eV$^{2}$, for 3~years
of operation) in the no-oscillation case. The discovery potential
is around 0.04 (3 standard deviations). The Double Chooz collaboration
plans to start taking data at the Chooz far/near site in 2007/2008.

Kaska is a Japanese collaboration aiming to start data taking end
of 2008. Kaska~\cite{Kaska} could be located close to the KAShiwazaki-KAriwa
nuclear power station (BWR, 24.3 GW$_{\textrm{th}}$). The plant is
composed of 7 cores divided into 2 clusters spread by 2~km. Thus,
two near detectors are mandatory (each at 400~m from a cluster).
The Kaska design is similar to the Double Chooz one\,: a 10~ton target
of Gd doped liquid scintillator and a $\gamma$-catcher region enclosed
in a double acrylic sphere, gamma shielding, a PMT supporting structure,
and a weak scintillating region acting as a muon veto. The systematic
error foreseen is between 0.5 and 1~\%. The sensitivity is expected
to be between $\sin^{2}(2\theta_{13})<0.017-0.027$ (90~\%~C.L.,
for 3~years of operation, depending on the true value of $\Delta m_{atm}^{2}$),
in the no-oscillation case. 

\subsubsection{Second generation experiments}

A second generation of experiments, with a large target mass, and
ultra low and very well understood backgrounds, has been proposed
in order to further improve the sensitivity to $\sin^{2}(2\theta_{13})$.
These experiments would be complementary to the lower limit's of sensitivity
of the Superbeam program which probes the ($\theta_{13}$,$\delta$-CP)
plane~\cite{Deltareactor,AutieroDeclais}. The goal is to reach a sensitivity $\sin^{2}(2\theta_{13})<0.01$
(90~\%~C.L.). 

The Daya Bay experiment~\cite{Dayabay} could be located in the Guang-Dong
Province, close to the Daya Bay nuclear power plant (PWR, 11.6~GW$_{\textrm{th}}$).
Since the power station has been built near a mountain, the plan is
to excavate horizontal tunnels and several halls. This scheme would
provide an excellent overburden for the detectors: 400~mwe at a distance
of 300~m from the cores, and 1200~mwe at a distance of about 1.5-2.0~km.
This provides flexibility to optimize the detector rooms location
according to the volatile best fitted $\Delta m²_{atm}$ value. The
particularity of this experiment is to use the so-called {}``movable
detector'' concept. The three detectors could be swapped for cross
calibration. The question of the systematic error introduced when
moving such sensitive machines remains open, however. 

The Braidwood experiment~\cite{Braidwood} could be located close
to the Braidwood twin nuclear station (BWR, 7.2~GW$_{\textrm{th}}$),
in Illinois. The area surrounding the power plant has a flat topology,
thus two 120~m deep shafts as well as two large detector rooms have
to be excavated. The overburden of 450~mwe would provide the same
background contribution in each detector. Since all civil construction
have to be realized, the detector locations could be optimized according
to the true value of~$\Delta m²_{atm}$. The plan is to have $\geq$1
near detector of~25-50 tons (fiducial mass) at 270~m in the near
shaft, and $\geq$2 far detectors identical to the near ones, at $\sim$1.8~km
in the far shaft. The detector design omits the $\gamma$-catcher
region, and thus differs significantly from the other proposals. As
in Daya Bay the detectors could be swapped for cross calibration,
but using a platform transporter as well as a high capacity crane
for the operation, instead of a railway system. 

The Angra experiment~\cite{Angra}, near the 6 GW$_{\textrm{th}}$
power station of Angra dos Reis, is going to focus on a high-luminosity
approach to provide a full energy spectrum measurement of the oscillation
signature. The far detector site could be located at 1.5~km from
the primary reactor core, under 700~m of granite (1700~mwe). The
detector would be a 500 ton fiducial volume of Gd loaded liquid scintillator.
The near detector could be either identical to the far detector at
300~m from the core (covered by 100~m of granite), or smaller than
the far detector (non-identical) but very near to the nuclear core.
If a luminosity of 6,000~GW$_{\textrm{th}}$-ton-year can be achieved,
the expected sensitivity is~$\sin^{2}(2\theta_{13})<0.007$.

\section{Neutrinos and society}

In the past, neutrino experiments have only been used for fundamental
research, but today, thanks to the extraordinary progress of the field,
e.g. the measurement of the oscillation parameters, neutrinos could
be useful for Society. We will see that reactor experiments could
play an important role in this new field in the near future.

\subsection{Non proliferation of nuclear weapons and reactor monitoring}

The International Atomic Energy Agency (IAEA) works with its Member
States to promote safe, secure and peaceful nuclear technologies.
One of its missions is to verify that safeguarded nuclear material
and activities are not used for military purposes. In a context of
international tension, neutrino detectors could help the IAEA to verify
the Treaty on the Non-Proliferation of Nuclear Weapons (NPT), signed
by 145~States around the world. 

A small neutrino detector located at a few tens of meters from a nuclear
core could monitor nuclear reactor cores non-intrusively, robustly
and automatically. Since the antineutrino spectra and relative yields
of fissioning isotopes~$^{235}$U, $^{238}$U, $^{239}$Pu, $^{241}$Pu
depend on the isotopic composition of the core, small changes in composition
could be observed without ever directly accessing the core itself.
Information from a modest-sized antineutrino detector, coupled with
the well-understood principles that govern the core's evolution in
time, can be used to determine whether the reactor is being operated
in an illegitimate way (see Figure~\ref{AntinueSpectra}). Furthermore,
such a detector can help to improve the reliability of the operation,
by providing an independent and accurate measurement, in real time,
of the thermal power and its reactivity at a level of a few percent.
The intention is to design an ``optimal'' monitoring detector by using
the experience obtained from neutrino physics experiments and feasibility
studies. 

Sands is a one cubic meter antineutrino detector located at 25~meters
from the core of the San Onofre reactor site in California~\cite{Sands}.
The detector has been operating for several months in an automatic
and non-intrusive fashion that demonstrates the principles of reactor
monitoring. Although the signal to noise ratio of the current design
is still less than two, it is possible to monitor the thermal power
at a level of a few percent in two weeks. At this stage of the work,
the study of the evolution of the fuel seems difficult, but this has
already been demonstrated by the Bugey and Rovno experiments~\cite{Rovno,Bugey}. 

Double Chooz will be a research detector with a very high sensitivity
to study neutrino oscillations. About 20,000~neutrino events per
year are expected in the far detector, but millions of events in the
near detector (between 100 and 200 m away from the cores). These huge
statistics could be exploited to help the IAEA in its Safeguards missions~\cite{DoubleChoozEU}.
The potential of neutrinos to detect various reactor diversion scenario's
can be tested with Double Chooz. A realistic reactor monitor is likely
to be somewhere between the two concepts presented above.

\subsection{Geophysics}

The total measured heat dissipation rate from the Earth lies between 30 to 40 TW.
Geo-neutrinos are the neutrinos and antineutrinos from the progenies of U, Th and K 
decays in the Earth, and Earth composition models suggest that about 20 TW originate from these isotope decays
(see for example \cite{Fiorentini} for more information on geo-neutrinos). U and Th end-point energies are above the threshold of inverse beta reactions on free protons, thus they can be detected as reactor antineutrinos. U and Th geoneutrinos could in principle be distinguished due to their different energy spectra, e.g. geo-neutrinos with energy $E>2.25$~MeV are produced only from the Uranium chain. In conclusion, antineutrinos could be used for determining the radiogenic contribution to the terrestrial heat flow and for discriminating among different models of the Earth's composition. 

Since 2002 the KamLAND experiment has looked for geo-neutrinos and has just published updated data~\cite{KamGeo05}. 
They have detected $4.5$ to $54.2$ geoneutrino events (90\,\% C.L.), whereas 19 events were predicted by the standard geophysical model. This provide, for the first time, a direct upper limit of 60 TW (99\,\% C.L.) for the radiogenic power of U and Th in the Earth~\cite{KamGeo05}.

In the meanwhile,
other projects for geo-neutrino detection are being planed. Borexino
at Gran Sasso will
benefit from the absence of nearby reactors. The LENA proposal~\cite{Lena}
envisages a 50~kton liquid scintillator detector at the Center for
Underground Physics in the Pyhas\"{a}lmi mine (Finland). Due to the
huge mass, it should collect several hundred events per year. In conclusion,
one can expect that within 10 to 20 years the geo-neutrino signal
from Uranium and Thorium will be measured at a few points on the globe. 

A natural nuclear fission reactor with a power output of 3 to 10~TW
at the center of the Earth has been proposed as the energy source
of the Earth's magnetic field~\cite{Herndon}. The proposal can be
directly tested by a massive liquid scintillator detector that can
detect the signature spectrum of antineutrinos from the geo-reactor
as well as the direction of the antineutrino source (Earth core).
However, the clarity of both types of measurements may be limited
by background from antineutrinos from surface power reactors. 





\end{document}

<x:/)€mÿÿÿÿSquare 

<x:€ÿÿÿÿÿÿÿÿ
Big Operators€µ€ÿÿÿÿSum€¸€
<x:€¹€ÿÿÿÿoint

<x:€·€ÿÿÿÿ	Coproduct
Font Si&zeerators
<x:ÿÿÿÿpm€–€
<x:€—€ÿÿÿÿtimes

<x:€™€ÿÿÿÿast€€ÿÿÿÿcdot
Font Si&ze€œ€
<x:
<x:€ÿÿÿÿotimes€¬€
<x:€­€ÿÿÿÿominus€°€ÿÿÿÿodot€¯€ÿÿÿÿoslash
<x:€ÿÿÿÿ€¿€ÿ
<x:€À€ÿÿÿÿ
big oti
ÈAøf<p[<ÈAhh< i<$ÈAj<Hk<0ÈA€˜<8™<<ÈA¨š<`›<HÈAМ<¨M<ig uplus
<x:ÿÿÿÿstar€ €
&Alignment¡€ÿÿÿÿsqcap€¢€ÿÿÿÿsqcup
<x:€ÿÿÿÿsetminus€¦€
<x:€¨€ÿÿÿÿbig triangle up€©€ÿÿÿÿbig triangle down€
<x:rc€²€ÿÿÿÿdagger
<x:ÿÿÿÿddagger€´€ÿÿÿÿamalg€ÿÿÿÿÿÿÿÿ
Comparison
<x:€Ì€ÿÿÿÿgeq€Ä€ÿÿÿÿp
Font Si&zeÿÿÿÿsucc
<x:
<x:€«€ÿÿÿÿtriangle righ
<x:€Û€ÿÿÿÿneq€Õ€ÿÿ
&MakeIndex€Ù€ÿÿÿÿapprox€Ú
<x:€å€ÿÿÿÿpropto
<x:ÿÿÿÿ€Æ€ÿÿÿÿll
Next Error€É€ÿÿÿÿ
sqsubsete
<x: